\documentclass[12pt]{article}
\usepackage[dvips]{graphics}
\def\R{{\bf R}}
\def\d{\partial}
\def\tr{{\rm tr}}

\begin{document}

\title{Manufacture of Gowdy spacetimes with spikes}

\author{Alan D. Rendall and Marsha Weaver\\ \\
Max-Planck-Institut f\"ur Gravitationsphysik\\ Am M\"uhlenberg 1\\
D-14476 Golm, Germany}

\maketitle

\begin{abstract}
In numerical studies of Gowdy spacetimes evidence has been found for the
development of localized features (\lq spikes\rq ) involving large gradients 
near the singularity. The rigorous mathematical results available up to now 
did not cover this kind of situation. In this work we show the existence of
large classes of Gowdy spacetimes exhibiting features of the kind 
discovered numerically. These spacetimes are constructed by applying
certain transformations to previously known spacetimes without spikes.
It is possible to control the behaviour of the Kretschmann scalar near
the singularity in detail. This curvature invariant is found to blow up in 
a way which is non-uniform near the spike in some cases. When this happens
it demonstrates that the spike is a geometrically invariant feature and
not an artefact of the choice of variables used to parametrize the metric.
We also identify another class of spikes which are artefacts. The spikes 
produced by our method are compared with the results of numerical and
heuristic analyses of the same situation. 
\end{abstract}

%\pacno{04.20.Dw,04.20.Ex}

\section{Introduction}

The question of the nature of spacetime singularities in general relativity
is far from being answered in general. The encouraging progress which has
been made on this problem in the past few years has mainly consisted of
obtaining detailed information about certain phenomena in more or less
special classes of spacetimes. The effects which have been studied
in a rigorous mathematical way include the decoupling of the evolution at 
different spatial points near the singularity (see \cite{andersson00}, 
\cite{kichenassamy98}, \cite{rendall00a} and references therein) and the 
monotone or oscillatory approach to the singularity, which has been 
investigated in the case of spatially homogeneous spacetimes 
(\cite{weaver00a}, \cite{ringstrom00a}, \cite{ringstrom00b}, 
\cite{rendall99a}, \cite{rendall00b}). In numerical work it
has been possible to study both together (\cite{berger98a}, \cite{weaver98a},
\cite{berger98b}). The first of these effects means that
an inhomogeneous model can be approximated near the singularity by a family 
of homogeneous models parametrized by the spatial coordinates and this is 
why the second is of wider importance. This choice of topics was stimulated 
by the pioneering ideas of Belinskii, Khalatnikov and Lifshitz (BKL) (see
\cite{bkl82} and references therein).

One class of spacetimes which has turned out to be particularly
useful in developing ideas about the nature of spacetime singularities
is that of Gowdy spacetimes. These do not show an infinite number of 
oscillations as the singularity is approached in the way that generic
spacetimes are supposed to do according to the BKL picture. Instead 
they appear to show a monotone behaviour near the singularity which is 
called velocity dominated (\cite{eardley72}, \cite{isenberg90}). Thus 
they provide a relatively simple arena for 
studying spacetime singularities. On the other hand, as will become clear 
in the following, the Gowdy spacetimes do display quite complicated 
dynamical behaviour near their initial singularities and it is probable 
that what we can learn from them will be useful in understanding  
more general classes of spacetimes.

The Gowdy spacetimes which have been studied most extensively are those
with spatial topology $T^3$ and only these will be discussed in the
following. Grubi\v{s}i\'c and Moncrief \cite{grubisic93} derived a consistent
formal expansion for Gowdy spacetimes near their singularities under
a certain assumption. For Gowdy spacetimes it is possible to consider 
a quantity $v$ called the velocity. (Since $v$ is a non-negative scalar 
quantity it would be more appropriate to call it speed, but we will
retain the usual terminology.) If velocity dominance holds this
quantity has a (space-dependent) limit as the singularity is 
approached. It is also possible to associate a sign to each spatial
point and thus define a real-valued function $k$ with the given 
magnitude and sign. This function will be called the asymptotic velocity. 
The Grubi\v{s}i\'c-Moncrief expansion is consistent provided $0<k<1$, 
which defines what is called the low velocity regime. For high velocity 
($k\ge 1$) this is no longer true in general. (Some remarks on the meaning 
of $k\le 0$ will be made later.) It was suggested in \cite{grubisic93} that 
in those cases where the expansion is not consistent the solution evolves 
in such a way that the velocity is eventually less than one close to the 
singularity.  It was also noticed that under an additional non-generic 
condition it is possible for the expansion to be consistent with high 
velocity. 

A little later Berger and Moncrief \cite{berger94a} observed
the development of large spatial derivatives near the singularity
in numerical calculations of Gowdy spacetimes. They called
these \lq spiky features\rq. We will abbreviate this to \lq spikes\rq.  
They presented evidence that these were genuine features of the
spacetimes and not numerical artifacts. Spikes were seen to be associated
with the situation where the non-generic condition of \cite{grubisic93} 
is satisfied 
at an isolated point. In this case it can happen that the velocity stays 
high at the special point as the singularity is approached while it is 
forced below one everywhere else. This gives an intuitive picture of
how under certain circumstances strongly spatially inhomogeneous behaviour 
may develop in Gowdy spacetimes as the singularity is approached.

In the special case of the polarized Gowdy spacetimes velocity dominance 
has been proved \cite{isenberg90}. There is also a theorem on the structure
of the initial singularity in those non-polarized vacuum models evolving 
from data which are close to those for a particular Kasner solution (with 
exponents $\frac{2}{3},\frac{2}{3},-\frac{1}{3}$) \cite{chrusciel91}.
The model solution has velocity $v$ identically zero. The discovery of spikes
in the evolution of general Gowdy models implies limits to the way in which
these results for particular subclasses could generalize.  

A complementary development to these theorems on the nature of the
singularity arising from given Cauchy data for a Gowdy spacetime is
the construction of large classes of solutions with singularities of a
particular type. This was carried out for the case of velocity dominated
singularities with asymptotic velocity satisfying $0<k<1$ by
Kichenassamy and Rendall \cite{kichenassamy98}. Solutions were 
constructed depending on four free functions, which is the same number 
as in the general solution. Intuitively this suggests that solutions 
corresponding to an open set of initial data are being obtained by this 
procedure, although it has not been possible to prove this 
yet. Solutions depending on three free functions were also constructed 
under the weaker restriction $k>0$. In  \cite{kichenassamy98} the free 
functions were required to be analytic, but this undesirable restriction 
was removed in \cite{rendall00a}. In the following the low velocity 
solutions of \cite{kichenassamy98} and \cite{rendall00a} will be used as 
a starting point to produce another family of solutions, depending on the 
same number of free functions, which exhibit spikes.

If one looks at numerically produced spikes (see e.g.\ figure 2 in
\cite{berger97a}) then it becomes clear that there are two different 
types and that within any one of these types the individual spikes
show a great similarity among themselves. This is a strong motivation
for finding a simple explanation, and the main result of this paper
is to do this in a mathematically rigorous way. The principal unknowns
in the Einstein equations for Gowdy metrics are real-valued functions 
$P$ and $Q$. In one case $P$ has a spike and $Q$ has a kind of 
discontinuity while in the other case $P$ has a spike of the same shape
but pointing in the opposite direction while $Q$ is smooth. For reasons
to be explained later, these will be called \lq false\rq\ and \lq true\rq\ 
spikes, respectively. A key feature is that in the first case geometric 
quantities such as curvature invariants show no irregular behaviour at 
the spike as the singularity is approached, while in the second case
they do.

The field equations in Gowdy spacetimes are known to be closely 
related to the geometry of the hyperbolic plane (see, for instance,
\cite{grubisic93}). The interpretation in terms of hyperbolic geometry leads 
rather easily to a transformation (inversion) which produces solutions
with false spikes from the smooth low velocity solutions of 
\cite{kichenassamy98}. More 
subtle is a further transformation (Gowdy-to-Ernst transformation) which 
converts solutions with false spikes into solutions with true spikes. It 
is related to the Kramer-Neugebauer transformation 
(\cite{kramer68}, \cite{breitenlohner87}) occurring in
the theory of stationary and axisymmetric spacetimes.

The paper is organized as follows. In section 2 the basic notations and
equations are introduced and the two transformations of key importance in
the following are defined. The third section introduces the concepts of
false and true spikes, shows how solutions displaying features of this kind
can be produced and gives asymptotic expansions for these solutions near the 
spikes. In the fourth section the analytical and numerical results are
compared. Section 5 relates what we have found here to the method
of consistent potentials and hence to the dynamics which produces these
solutions. Finally, section 6 discusses some further developments of the
results.

\section{The Gowdy spacetimes}\label{gowdysec}

The Gowdy spacetimes on $T^3$ are solutions of the vacuum Einstein equations 
which can be written in the form
\begin{equation}
\label{metric}
g = - e^{(\lambda - 3 \tau) / 2} d \tau^2 +
e^{(\lambda + \tau) / 2} d x^2 +
e^{P - \tau} (dy + Q \, dz)^2 + e^{-P-\tau} dz^2
\end{equation}
They admit $U(1)\times U(1)$ as an isometry group, acting by translation of
the periodic coordinates $y$ and $z$. The notation here differs somewhat 
from that used in  \cite{kichenassamy98}. The quantities $(P,Q,\tau)$ 
occurring here are related to the quantities $(X,Z,t)$ of 
\cite{kichenassamy98} by 
$$t = e^{-\tau} \hspace*{20pt} P = -Z \hspace*{20pt}Q = X $$
The coordinate $t$, and hence the coodinate $\tau$, is determined in an
invariant way by the geometry: $t$ is proportional to the area of
the the orbits of the symmetry group. 
The functions $P$ and $Q$ depend on $\tau$ and $x$. For a Gowdy spacetime
on $T^3$ they should be periodic in $x$, but most of this paper is concerned
with constructions which are local in $x$. Information about the initial 
singularity, which is at $t=0$, is encoded here in the asymptotic behaviour 
of the metric as $\tau\to\infty$. The vacuum Einstein equations imply that 
the functions $P$ and $Q$ satisfy
\begin{eqnarray}
\label{gowdyeq}
P_{\tau \tau} & = & e^{2 P} Q_\tau^2 + e^{-2 \tau}
(P_{x x} - e^{2 P} Q_x^2) \nonumber \\
Q_{\tau \tau} & = & -2 P_\tau Q_\tau + e^{-2\tau}
(Q_{x x} + 2 P_x Q_x)
\end{eqnarray}
and that $\lambda$ satisfies
\begin{eqnarray}
\lambda_\tau & = & -(P_\tau^2 + e^{2P} Q_\tau^2)
- e^{-2 \tau} ( P_x^2 + e^{2 P} Q_x^2) \nonumber \\
\lambda_x & = & -2(P_\tau P_x + e^{2P} Q_\tau Q_x)
\end{eqnarray}
These equations determine $\lambda$ up to a constant. When the above 
equations are satisfied $P$, $Q$ and $\lambda$ define a solution of
the vacuum Einstein equations via (\ref{metric}).
 
Suppose that a solution $(P_0,Q_0)$ of (\ref{gowdyeq}) is given. Define
functions $(P_1,Q_1)=I(P_0,Q_0)$ by
\begin{equation}
\label{inversion}
e^{-P_1} = {e^{-P_0} \over Q_0^2 + e^{-2P_0}} \hspace*{40pt}
Q_1 = {Q_0 \over Q_0^2 + e^{-2P_0}}
\end{equation}
Then $(P_1,Q_1)$ also satisfy (\ref{gowdyeq}), as can be checked by
direct calculation. The transformation 
$(P_0,Q_0) \stackrel{I}{\longrightarrow} (P_1,Q_1)$
has been denoted by $I$, since it has a geometric interpretation in terms
of an inversion in the hyperbolic plane. This will now be explained in
some detail.

The description of Gowdy spacetimes in terms of the three real-valued 
functions $P$, $Q$ and $\lambda$ of $\tau$ and $x$ can be replaced by one 
with greater invariance which leads to useful insights. For this purpose
it is useful to think of these as functions of $t$ and $x$ which, with
a slight abuse of notation, will be denoted by the same letters. Let $M$ be 
the manifold with coordinates $t$ and $x$, which is just an open subset
of $\R^2$. Let the flat metric $-dt^2+dx^2$ on $M$ be denoted by $f$.
Define a mapping from $M$ to $\R^2$ by the pair $U=(P,Q)$. Let $(N,h)$ be
$\R^2$ with the metric $h$ defined by $dP^2+e^{2P}dQ^2$. The derivative 
$dU$ is a section of the tensor product of the cotangent bundle of $M$
with the pull-back via $U$ of the tangent bundle of $N$ to $M$. There
is a natural connection defined on this bundle by means of the 
Levi-Civita connections associated to the metrics $f$ and $h$. This 
defines a covariant derivative, call it $D$. The Gowdy equations for 
$P$ and $Q$ take the form $\tr_f(DdU)-t^{-1}U_t=0$. This a slight 
generalization of the equation for a wave map (also known as a hyperbolic 
harmonic map or nonlinear $\sigma$-model) which is $\tr_f(DdU)=0$. The 
functions of $P$ and
$Q$ occurring in the equations for $\lambda$ are also expressible in
terms of $h$. They are just $h(U_t,U_t)+h(U_x,U_x)$ and $h(U_t,U_x)$.
It follows that all the Gowdy field equations can be expressed in a
way which is covariant with respect to $(N,h)$. An advantage of this is
that applying any isometry of $h$ to a solution of the Gowdy equations
will give a new solution. It can be shown that the metric corresponding
to the functions $(P,Q,\lambda)$ after transformation is isometric to
the original one, the transformation being effected by a linear change
of the variables $y$ and $z$. This will not be proved here, since we 
only need it in one special case (inversion), where it can be checked 
directly. The fact that the two metrics are isometric means that quantities 
such as invariants of the spacetime curvature are identical for the 
original solution and the transformed solution.

Another way of looking at the invariance properties of the equations and
their relation to wave maps is via a Lagrangian formulation. A wave map 
from two-dimensional Minkowski space to the hyperbolic plane is the 
Euler-Lagrange equation corresponding to the action 
\begin{eqnarray}
&&\int f^{ij}\d_i U^A\d_j U^B h_{AB} dV(f)    \nonumber \\
&=&\int ((-P_t^2+P_x^2)+e^{2P}(-Q_t^2+Q_x^2)) dt dx  
\end{eqnarray}
The Gowdy equations are the Euler-Lagrange equations for the analogous
action
\begin{equation}
\int ( (-P_\tau^2 + e^{-2 \tau} P_x^2) + e^{2P}(-Q_\tau^2
+ e^{-2 \tau} Q_x^2) ) d\tau dx
\end{equation}

One isometry of the metric $h$ is given by the transformation 
(\ref{inversion}).
This was written down in a slightly different notation in 
\cite{kichenassamy98}. Its geometrical origin is as follows.
Define $Y=e^{-P}$. The metric $h$ is that of hyperbolic space and 
the coordinates $(Q,Y)$ give the well-known upper half-space   
representation of that space on the region $Y>0$. In that representation
the transformation introduced above becomes inversion in the unit circle
and it interchanges the origin with the point at infinity. For this
reason this transformation is referred to as inversion $I$ in the 
following. It is induced by the linear transformation of coordinates
which interchanges $y$ and $z$. Another representation of the hyperbolic
space is given by the Poincar\'e disk model, 
where the points of the boundary of the half-space model are all on an 
equal footing with the point at infinity. From this point of view it is 
clear that the coordinates $(P,Q)$ break the symmetry of the hyperbolic 
plane by singling out a point of its ideal boundary as the point at 
infinity. The inversion $I$ changes this choice.

Next another transformation which generates new solutions from old will
be introduced. We know of no simple geometric interpretation such as that
given for $I$ above. In contrast to $I$, which transforms one coordinate
representation of a metric into another representation of the same metric,
the transformation which will be considered now produces a genuinely new
metric (i.e.\ one which is not isometric to the original one). This
transformation was suggested by the fact, already discussed in detail
in \cite{chrusciel90}, that a given Gowdy spacetime defines two pairs of
functions which satisfy identical equations. These come from the Gowdy
and Ernst methods of reduction with respect to the Killing vectors of 
the spacetime and so we will call this transformation GE for 
Gowdy-to-Ernst. If $(P_1,Q_1)$ is a solution of the Gowdy equations 
then the transformation 
$(P_1,Q_1) \stackrel{GE}{\longrightarrow} (P_2,Q_2)$  is defined by
\begin{equation}
\label{gowernst}
P_2 = -P_1 + \tau \hspace*{20pt}
Q_{2 \tau} = -e^{2(P_1 - \tau)} Q_{1 x} \hspace*{20pt}
Q_{2 x} = -e^{2P_1} Q_{1 \tau}
\end{equation}
In fact, $Q_2$ is only defined up to an additive constant by these relations
and it is necessary to be careful about specifying the constant when 
applying the transformation for a specific purpose. The integrability 
condition only assures the existence of a solution locally in $x$ and
in most of this paper we will restrict to that situation. See, however,
the comments on spikes in (global) Gowdy spacetimes in section 
\ref{further}. The transformation $GE$ is such that if $(P_1,Q_1)$
satisfy the Gowdy equations~(\ref{gowdyeq}) then
$(P_2,Q_2)$ do so too.  That the integrability condition for
$Q_{2 \tau}$ and $Q_{2 x}$ is satisfied can be verified directly
by calculating the $x$-derivative of the right hand side of the
expression for $Q_{2 \tau}$ and
the $\tau$-derivative of the right hand side of the
expression for $Q_{2 x}$ and seeing that they agree if
$P_1$ and $Q_1$ satisfy the Gowdy equations. There is a close relationship
between this transformation and the Kramer-Neugebauer transformation for
stationary axisymmetric spacetimes (see \cite{kramer68}, 
\cite{breitenlohner87}). They differ by no more than signs related to
the different character of the Killing vectors (both spacelike in the
one case, one timelike and one spacelike in the other case).

Since the relations (\ref{gowernst}) only fix $Q_2$ up to a constant,
there is in fact a whole one-parameter family of solutions produced
by this procedure which differ by translations in the $Q$-direction. On the
other hand, changing $Q_1$ by a constant does not change $Q_2$.

\section{Construction of solutions with spikes}\label{manufacture}

The central procedure of this paper is the manufacture of solutions of
the Gowdy equations with spikes out of solutions without spikes by
means of the transformations $I$ and GE introduced in the last section.
The source of solutions without spikes which we draw on is a class
of solutions constructed in \cite{rendall00a} and so we begin by
describing those solutions in terms of the variables used here. Let 
$\epsilon$ be a positive constant and let $k(x)$, $X_0(x)$, $\phi(x)$
and $\psi(x)$ be periodic $C^\infty$ functions, i.e.\ smooth functions 
defined on $S^1$. Suppose that $\epsilon<2k(x)$, $\epsilon<2(1-k(x))$ and 
$0<k(x)<1$. Then there exist unique $C^\infty$ functions $u(\tau,x)$ and
$v(\tau,x)$, periodic in $x$, which converge to zero in the limit that
$\tau \rightarrow \infty$ such that the functions
$P(\tau,x)$ and $Q(\tau,x)$,
\begin{equation}
\label{expansion0}
P = k \tau - \varphi - e^{- \epsilon \tau} \, u \hspace*{40pt}
Q = X_0 + e^{-2 k \tau} \, ( \psi + v )
\end{equation}
satisfy the equations (\ref{gowdyeq}). A similar statement holds locally.
If the input functions $k$, $X_0$, $\phi$ and $\psi$ are defined for $x$
belonging to some open interval, then a unique solution as above is 
obtained for $x$ belonging to some slightly smaller open interval. The 
function $k$ coincides with the asymptotic velocity mentioned in the
introduction and is positive in this case. Note
that statements of this kind were originally proved in \cite{kichenassamy98}
under the additional hypothesis of analyticity of the input functions.
We will refer to these solutions (whether analytic or not) as the low 
velocity KR (Kichenassamy-Rendall) solutions.

It follows from what has been said about $P$ and $Q$ above that 
\begin{equation}
P = k \tau - \varphi + o(1)\hspace*{40pt}
Q = X_0 + e^{-2 k \tau} \, \psi + o(e^{-2k\tau})
\end{equation}
It is also important for the following to know that these asymptotic
expansions can be differentiated term by term in an appropriate sense.
This is a consequence of the fact that the functions $u$ and $v$ are 
regular in the sense of Definition 1 in \cite{rendall00a}. It can be
concluded that 
\begin{equation}
\d_x^j P = \d_x^j k \tau - \d_x^j\varphi + o(1)
\end{equation} 
for all $j\ge 0$. For $Q$ we have the slightly more complicated statement
that
\begin{equation}
\d_x^j Q = \d_x^j X_0 + \d_x^j(e^{-2 k \tau} \, \psi) + 
o(\tau^j e^{-2k\tau})
\end{equation}
In this relation it is also possible to replace $\d_x^j(e^{-2 k \tau}\psi)$
by $\d_x^j k (-2\tau)^j e^{-2 k \tau}\psi$. Estimates for time derivatives
can be obtained by substituting these expressions into
equations~(\ref{gowdyeq}). (To be
precise, this is true for time derivatives of order at least two. The 
statements for the first order time derivatives are direct consequences of
the theorems of \cite{rendall00a}, which were proved using a 
first order formulation of the equations, where the time derivatives
of the basic variables are included as unknowns.) There results the estimate
\begin{equation}
\d_\tau^l\d_x^j P = \d_\tau^l\d_x^j (k \tau) + o(1)
\end{equation}
for all $j\ge 0$ and $l > 0$. Of course the first term on the right hand
side vanishes as soon as $l>1$. Similarly
\begin{equation}
\d_\tau^l\d_x^j Q = \d_\tau^l\d_x^j(e^{-2 k \tau} \, \psi) + 
o(\tau^j e^{-2k\tau})
\end{equation}

Consider now a low velocity KR solution $(P_0,Q_0)$ such that the 
corresponding function
$X_0$ vanishes at an isolated point $x_{spike}$. Applying the transformation
$I$ produces a new solution $(P_1,Q_1)$ of the Gowdy equations according to
equations (\ref{inversion}). The intuitive idea is that the new solution
has a spike at $x_{spike}$, a statement which will be made precise later.
Applying the transformation $I$ to (\ref{expansion0}) with $(P,Q)$ replaced
by $(P_0,Q_0)$, one finds that
\begin{eqnarray}
\label{falseeq}
P_1 & = & k \tau - \varphi - e^{-\epsilon \tau} u +
\ln\{ [ X_0 + e^{-2 k \tau} ( \psi + v ) ]^2
+ e^{-2 k \tau} e^{2 \varphi} e^{2 e^{-\epsilon \tau} u} \}
\nonumber \\
Q_1 & = & { X_0 + e^{-2 k \tau} ( \psi + v ) \over
[ X_0 + e^{-2 k \tau} ( \psi + v ) ]^2
+ e^{-2 k \tau} e^{2 \varphi} e^{2 e^{-\epsilon \tau} u} }
\end{eqnarray}
The aim now is to investigate the qualitative behaviour of this solution
in the limit $\tau\to\infty$. This will be done by fixing a point $x$ 
and expanding in $\tau$ in a suitable way. It turns out that the result
is quite different depending whether $x$ is equal to $x_{spike}$ or not.
At $x_{spike}$ one finds that
\begin{eqnarray}
P_1 & = & -k \tau + \varphi + e^{-\epsilon \tau} \tilde{u}_a   
\nonumber \\
Q_1 & = & e^{-2 \varphi}  ( \psi + v ) +
e^{-\epsilon \tau} \tilde{v}_a
\end{eqnarray}
for some functions $\tilde u_a$ and $\tilde v_a$ with 
$\lim_{\tau \rightarrow \infty} \tilde{u}_a = 0$
and $\lim_{\tau \rightarrow \infty} \tilde{v}_a = 0$.
Elsewhere one finds that
\begin{eqnarray}
P_1 & = & k \tau - \varphi + \ln X_0^2
+ e^{-\epsilon \tau} \tilde{u}_b
\nonumber \\
Q_1 & = & {1 \over X_0} + e^{-2 k \tau}
(-{\psi \over X_0^2} - {e^{2 \varphi} \over X_0^3} +
\tilde{v}_b)
\end{eqnarray}
for some functions $\tilde u_b$ and $\tilde v_b$
with $\lim_{\tau \rightarrow \infty} \tilde{u}_b = 0$
and $\lim_{\tau \rightarrow \infty} \tilde{v}_b = 0$. 
At each spatial point the asymptotic behaviour in time is of a similar 
form to that of the KR solutions. However the coefficients of the
leading terms behave in a very non-uniform way as a function of $x$
close to the spike. The coefficient corresponding to $k(x)$, the 
asymptotic velocity, has a
discontinuity at $x_{spike}$ while the coefficient corresponding to
$X_0$ is unbounded near that point. This means that the spatial 
derivatives of the functions $P$ and $Q$ close to the point $x_{spike}$
grow much faster as $\tau\to\infty$ than is the case for the KR solutions.
This behaviour shows up as spikes in numerical calculations. Since it
is obviously an effect of the parametrization of the metric and not
a geometrical one, we refer to this kind of point as a \lq false spike\rq.
Note that at a false spike the asymptotic velocity is negative. This shows
that the sign of $k$ has no absolute geometric significance in general.

Applying the transformation GE to $(P_1,Q_1)$ gives a new solution
$(P_2,Q_2)$. It also has a spike at the point $x_{spike}$ and, as will
be proved later, curvature invariants show highly non-uniform behaviour
there. This shows in particular that in this case there has been a 
change in the geometry and so we refer to this kind of point as a 
\lq true spike\rq. In this case one finds that
\begin{eqnarray}
\label{trueeq}
P_2 & = & (1 - k) \tau + \varphi + e^{-\epsilon \tau} u -
\ln\{ [ X_0 + e^{-2 k \tau} ( \psi + v ) ]^2
+ e^{-2 k \tau} e^{2 \varphi} e^{2 e^{-\epsilon \tau} u} \}
\nonumber \\
Q_{2 \tau} & = & e^{-2 \tau} \{ -X_{0 x}
- e^{-2 k \tau} ( \psi_x + v_x ) + \nonumber \\
& & [ e^{2 k \tau} X_{0 x} + \psi_x + v_x
-2 k_x \tau ( \psi + v ) ]
[ X_0 + e^{-2 k \tau} ( \psi + v ) ]^2
e^{-2 \varphi} e^{-2 e^{-\epsilon \tau} u}
\nonumber \\ & & -2 k_x \tau X_0 +
2 [ X_0 + e^{-2 k \tau} ( \psi + v ) ]
[\varphi_x + e^{-\epsilon \tau} u_x] \} \nonumber \\
Q_{2 x} & = & - e^{-2 k \tau} v_\tau + ( - 2 k ( \psi + v )
+ v_\tau ) [ X_0 + e^{-2 k \tau} ( \psi + v ) ]^2
e^{-2 \varphi} e^{-2 e^{-\epsilon \tau} u}
\nonumber \\ & & - 2 k X_0 +
2 X_0 e^{-\epsilon \tau} ( u_\tau - \epsilon u )
+ 2 e^{-(2 k + \epsilon) \tau} ( \psi + v )
( u_\tau - \epsilon u )
\end{eqnarray}
At $x_{spike}$ one finds that
\begin{eqnarray}
P_2 & = & (1 + k) \tau - \varphi + e^{-\epsilon \tau} \tilde{u}_c
\nonumber \\
Q_{2 \tau} & = & -e^{-2 \tau} X_{0 x} +
e^{-2(1+k) \tau} ( -\psi_x + X_{0 x} \psi^2 e^{-2 \varphi}
+ 2 \psi \varphi_x + \hat{v}_c ) \nonumber \\
Q_{2 x} & = & - e^{-2 k \tau} v_\tau +
2 e^{-(2 k + \epsilon) \tau} (\psi+v) ( u_\tau - \epsilon u )
\nonumber \\ & & -2 k e^{-4 k \tau} \psi^3 e^{-2 \varphi}
+ e^{-4 k \tau} \check{v}_c
\end{eqnarray}
with $\lim_{\tau \rightarrow \infty} \tilde{u}_c = 0$,
$\lim_{\tau \rightarrow \infty} \hat{v}_c = 0$ and
$\lim_{\tau \rightarrow \infty} \check{v}_c = 0$.
Elsewhere one finds that
\begin{eqnarray}
P_2 & = & (1 - k) \tau + \varphi - \ln X_0^2
+ e^{-\epsilon \tau} \tilde{u}_d
\nonumber \\
Q_{2 \tau} & = & e^{-2(1-k) \tau} X_0^2 X_{0 x}
e^{-2 \varphi} e^{-2 e^{-\epsilon \tau} u} +
e^{-2 \tau} \{ -X_{0 x} + 2 X_0 X_{0 x} \psi e^{-2 \varphi}
\nonumber \\ & & + X_0^2 ( \psi_x - 2 k_x \tau \psi )
e^{-2 \varphi} - 2 k_x \tau X_0 + 2 X_0 \varphi_x +
\hat{v}_d \} \\
Q_{2 x} & = & -2 k X_0 +(-2 k ( \psi + v ) + v_\tau)
X_0^2 e^{-2 \varphi}
e^{-2 e^{-\epsilon \tau} u} \nonumber \\ & &
-4 k e^{-2 k \tau} \psi^2 X_0 e^{-2 \varphi}
+ 2 X_0 e^{-\epsilon \tau} ( u_\tau - \epsilon u)
+ e^{-2 k \tau} \check{v}_d
\end{eqnarray}
with $\lim_{\tau \rightarrow \infty} \tilde{u}_d = 0$,
$\lim_{\tau \rightarrow \infty} \hat{v}_d = 0$
and $\lim_{\tau \rightarrow \infty} \check{v}_d = 0$.

While the pointwise expansions which have been obtained so far contain 
much valuable information, it is also useful to derive 
uniform expansions of the solutions on suitable regions. This will
only be done for the true spikes, which are the main focus of interest
here. For a true spike $Q_2$ behaves in a regular way and so the point 
$x_{spike}$ does not have to be considered separately. It is possible 
to get a uniform asymptotic expansion for $\tau$ large and $x$ in a
neighbourhood of $x_{spike}$. The expansions for $P$ and $Q$ and their
derivatives imply corresponding expansions for $Q_{2\tau}$, $Q_{2x}$ 
and their derivatives. An expansion for $Q_2$ itself can be obtained 
straightforwardly by integration. All these expansions are of the same
form as those of the starting low velocity solution.

In the case of $P_2$ it is necessary to distinguish between two different
regions in order to get uniform expansions. Let $k_*$ be a positive number. 
The interesting case is where $k_*$ is slightly greater or slightly less 
than the value of $k$ at $x_{spike}$, and hence has the same property for 
$x$ close to $x_{spike}$. Let $\sigma(\tau,x)=e^{k_*\tau}|x-x_{spike}|$. 
Intuitively the function $\sigma$ is a measure of the relative influence of 
the spike and the part of the boundary with low velocity asymptotics. We 
will see that the spike dominates when $\sigma$ is small while the low 
velocity behaviour dominates when $\sigma$ is large. 

For convenience of notation, let $\delta x=|x-x_{spike}|$. Consider a
region $R_1$ where $\tau$ is sufficiently large, $\delta x$ sufficiently
small and $\sigma\ge 1$. We also make the genericity assumption that 
$|X_{0x}(x_{spike})|=\eta > 0$. Given $0<\alpha<1$ it can be assumed by
choosing $R_1$ to make $\delta x$ sufficiently small, that on the region
$R_1$ we have $|X_0|\ge \alpha\eta\delta x$ and
\begin{equation}
|X_0|^{-1}\le \alpha^{-1}\eta^{-1}e^{k_*\tau}
\end{equation} 
Now choose $k_*<k(x_{spike})$. Then
\begin{equation}
[X_0+e^{-2k\tau}(\psi+v)]^2+e^{-2k\tau}e^{2\phi}e^{2e^{-\epsilon\tau}u}
=X_0^2[1+O(e^{-2(k-k_*)\tau})]
\end{equation}
The logarithm of this expression is then equal to the logarithm of $X_0^2$
up to a remainder of which is $O(e^{-2(k-k_*)\tau})$. In this way it can
be seen that $P_2$ has an expansion on the region $R_1$ which is of the
same form as that for the starting function $P$. In particular this
expansion can be differentiated term by term.

Next consider a region $R_2$ where $\tau$ is sufficiently large, $\delta$
is sufficiently small and $\sigma\le 1$. Then we can carry out a very 
similar procedure. This time we choose an arbitrary $\alpha>1$ and
$k_*>k(x_{spike})$. It is convenient to rewrite the expression for $P_2$ as
\begin{equation}
P_2 = (1 + k) \tau - \varphi - e^{-\epsilon \tau} u -
\ln\{ 1 + [ e^{k \tau} X_0 +  e^{-k\tau}( \psi + v ) ]^2
e^{-2 \varphi} e^{-2 e^{-\epsilon \tau} u} \}
\end{equation}
On the region $R_2$ an estimate of the form 
$e^{k\tau}|X_0|\le \alpha\eta e^{(k-k_*)\tau}$ holds. Thus the logarithm
in the last expression for $P_2$ is $O( e^{2(k-k_*)\tau})$ on $R_2$. We
obtain a uniform asymptotic expansion for the solution on the region 
$R_2$ which can be differentiated term by term. This expansion is of
the same form as those for high velocity KR solutions.

The significance of the above observations concerning uniform asymptotic
expansions is that in order to determine the behaviour of curvature
invariants in the regions $R_1$ and $R_2$ it is sufficient to use the 
expressions already derived in \cite{kichenassamy98}, since all the
assumptions needed to do that computation are also satisfied on the
regions $R_1$ and $R_2$ in the present case, although for rather
different reasons. In particular the information about what happens
as a point of the boundary is approached with $x$ a constant other
than $x_{spike}$ can be read off from the asymptotics in the region 
$R_1$ while the information about what happens when the boundary 
is approached along $x=x_{spike}$ is given by the asymptotics in
the region $R_2$. It follows that the Kretschmann scalar 
$K=R^{\alpha\beta\gamma\delta}R_{\alpha\beta\gamma\delta}$ is 
given up to terms of lower order by 
$K=t^{-(\hat k^2+3)}B(\hat k)$ where 
$B(\hat k)=64(\hat k-1)^2(\hat k+1)^2(\hat k^2+3)^{-3}$ and
$\hat k$ is the asymptotic velocity at the given value of $x$. For
$x\ne x_{spike}$ this is $\hat k=1-k(x)$ while $\hat k(x_{spike})=
1+k(x_{spike})$. For the solutions under consideration $B(\hat k)$
never vanishes. Thus the Kretschmann scalar blows up like a 
more negative power of the Gowdy time $t$ at the spike than 
elsewhere. Since the Gowdy time is fixed invariantly by the spacetime
geometry, this provides clear evidence that the non-uniformity near a 
true spike is a feature of the geometry and cannot be removed by a 
reparametrization as can be done in the case of the false spike. Note 
that there is a uniform lower bound for the rate of blow-up of $K$ in 
a neighbourhood of the spike and so there can be no violation of cosmic 
censorship there. 

The mean curvature blows up like $t^{-(\hat k^2+3)/4}$. This shows 
that the singularity is a crushing singularity even in the presence
of a true spike. If the Kretschmann scalar is expressed in terms of
the mean curvature then the same power is obtained everywhere.
However the constant of proportionality changes discontinuously as
a function of $x$ as the spike is passed.

In \cite{hern99} Hern investigated the behaviour of curvature invariants
in dynamical numerical calculations of Gowdy spacetimes. Our results
are consistent with his and in some cases help to explain them. He 
investigated more invariants and more aspects of their behaviour than 
we have done. The approach used here should also suffice to obtain
more detailed information about more curvature invariants in the spikes
we manufacture, if desired.

\section{Comparison with numerical results}

In the previous section we produced solutions of the Gowdy equations
where the functions $P$ and $Q$ behave in an inhomogeneous way near 
the initial singularity. The motivation for this was to provide an
explanation and rigorous confirmation of various features which had
been observed in numerical calculations. In this section the analytical
and numerical results will be compared so as to show their similarities.

As already mentioned in the introduction, looking directly at the results 
of the numerical calculations leads to the conclusion that near the 
singularity, and for generic initial data, there are a finite number of 
points where $P$ has a spike and that these points can be classified into
two types according to whether the spike in $P$ points upwards or downwards
and whether $Q$ is smooth or also exhibits inhomogeneous behaviour. We
claim that the `false' and `true' spikes constructed in section 3
reproduce the qualitative features observed numerically.

\begin{figure}
\includegraphics{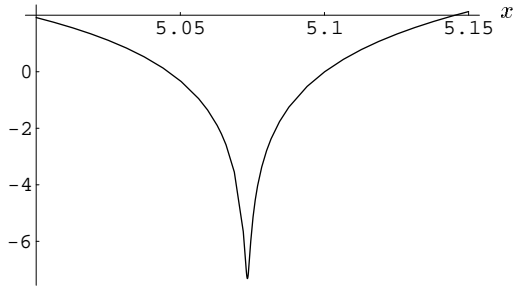}
\caption{$P_1$ in a neighbourhood of $x_{spike}$ at small $t$.}
\label{falsep}
\end{figure}

\begin{figure}
\includegraphics{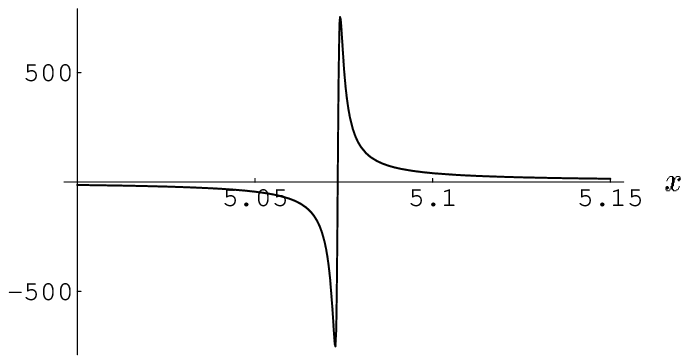}
\caption{$Q_1$ in a neighbourhood of $x_{spike}$ at small $t$.}
\label{falseq}
\end{figure}

\begin{figure}
\includegraphics{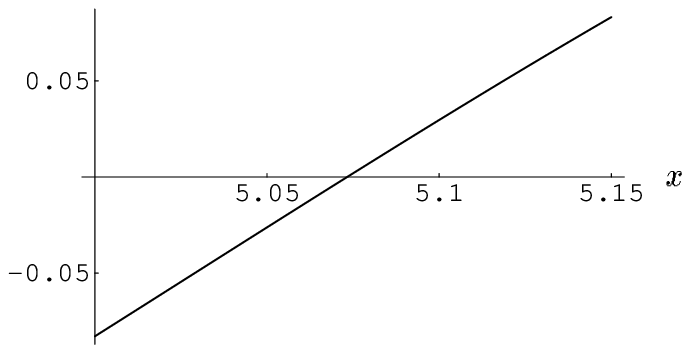}
\caption{$\pi_{Q 1}$ in a neighbourhood of $x_{spike}$ at small $t$.}
\label{falsepiq}
\end{figure}

Consider first false spikes. The coefficient of the leading term in the
expansion of the function $P_1$ has a discontinuous limit at the 
singularity. The asymptotic velocity jumps. At the spike it is between 
$0$ and $-1$. At fixed time we see the
influence of the expression $2X_0$ a little away from the spike. Under
the generic assumption that $X_{0x}(x_{spike})\ne 0$ this behaves like
a constant times $\log |x-x_{spike}|$ and accounts for the shape of the
flanks of the spike. The spike points downwards. The coefficient of the
leading term in the expansion for $Q_1$ is unbounded near the spike and at
fixed time behaves like a constant times $(x-x_{spike})^{-1}$ (assuming 
the genericity condition). This leads to a growing jump in $Q$ between 
positive and negative values as the singularity is approached. Another
quantity which has been studied numerically is the momentum conjugate
to $Q$ in a Hamiltonian formulation, $\pi_Q = e^{2P} Q_\tau$.
Calculation of $\pi_{Q 1}$ from $P_1$ and $Q_1$ shows that it tends to
zero as $t\to 0$ at $x_{spike}$.  These features are observed in numerical
calculations of the dynamics. To compare the two, we present pictures of
the results of applying the transformation $I$ to a low velocity KR solution
in figures~\ref{falsep}--\ref{falsepiq}.  For purpose of illustration
we choose periodic functions (with period $2 \pi$) for $k$, $X_0$,
$\varphi$ and $\psi$, and ignore the decaying unknowns $u$ and $v$.
The salient features of the spikes do not depend on the particular form
of the data, but to obtain a spike $X_0$ must vanish at an isolated point.
\begin{figure}
\includegraphics{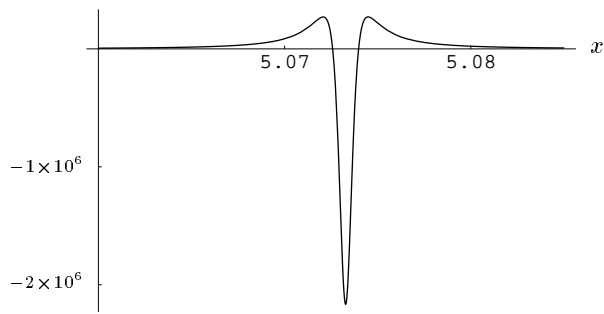}
\caption{$\pi_{Q_2}$ in a neighbourhood of $x_{spike}$ at small $t$.}
\label{truepiq}
\end{figure}

Consider next true spikes. Applying the GE transformation to the false
spike shown in figures~\ref{falsep}--\ref{falsepiq}, we get a true spike.
Since $P_2$ is obtained from $P_1$ by reflection (about the $x$-axis)
and translation (in the vertical direction) it is obvious that the profile
has the same shape as that of the false spike shown in figure~\ref{falsep},
except that in this case the spike points upwards instead of downwards.
On the other hand $Q_2$ tends to a continuous limit as $t\to 0$ and its
spatial derivative tends to zero at $x_{spike}$.  Since
$Q_{2 x} = - \pi_{Q_1}$, its graph is obtained from figure~\ref{falsepiq}
by reflection about the $x$-axis.  The most dramatic behaviour at a true
spike is exhibited by $\pi_Q$, as can be seen in figure~\ref{truepiq}.
The shape of $\pi_Q$ at a true spike can be understood from the
identity $\pi_{Q 2} = - Q_{1 x}$.  Figures~\ref{falsep}-\ref{truepiq},
along with $P_2$ and $Q_{2 x}$ as obtained from figures~\ref{falsep}
and \ref{falsepiq}, should be compared with figure 2 in \cite{berger97a}
and figures 2 and 3 in \cite{berger97b}.

\section{Comparison with the method of consistent potentials}\label{MCP}

Numerical simulations of $T^3$ Gowdy spacetimes consistently
show that the speed of the solution in hyperbolic space,
\begin{equation}
\label{velocity}
v = \sqrt{P^2_\tau + e^{2P} Q^2_\tau}
\end{equation}
is, asymptotically, less than one almost everywhere \cite{berger97a}.
Furthermore, the numerical simulations show that $v<1$ is obtained
through a series of steps.  On each step $v$ is essentially constant
in time.  From one step to the next, $v$ goes to approximately $|2 - v|$,
until $v < 1$ on the final step.  In the absence of spikes, on each step
the numerical solution is approximated well by a geodesic loop solution
\cite{grubisic93}, in which the solution follows a geodesic in hyperbolic
space at each spatial point.  A geodesic loop solution is a one parameter
family of Kasner solutions.  Thus the steps are known as Kasner Epochs.

In this setting the one parameter family of Kasners occurs as a solution
to the following differential equations for $P$ and $Q$.
\begin{eqnarray}
\label{gowdyinkasnerepoch}
P_{\tau \tau} & = & e^{2 P} Q_\tau^2 \nonumber \\
Q_{\tau \tau} & = & -2 P_\tau Q_\tau
\end{eqnarray}
The numerical simulations show that, during a Kasner Epoch, the terms on
the right hand side of the evolution equations~(2) which are absent from
the right hand side of equations~(\ref{gowdyinkasnerepoch}) are small.
However, unless $v<1$ or some non-generic condition is satisfied, one
of the absent terms eventually grows.  On the other hand, if $v<1$ the
numerical simulations show that the absent terms are all decaying.  The
Kasner solutions can be written as follows, with constants $\alpha>0$,
$\beta \geq 0$,  $\zeta$ and $\xi$ \cite{berger97a}.
\begin{eqnarray}
\label{kasner}
P & = & \ln \alpha + \ln(e^{-\beta \, \tau}
+ \zeta^2 e^{\beta \, \tau} ) \nonumber \\
Q & = & \xi - { \zeta e^{ 2 \beta \tau} \over
\alpha ( 1 + \zeta^2 e^{2 \beta \tau})}
\end{eqnarray}
A one parameter family of Kasners is
obtained by allowing $\alpha$, $\beta$, $\zeta$ and
$\xi$ to depend on $x$.  Calculation shows that the speed
in hyperbolic space for $(P,Q)$ as in equations~(\ref{kasner})
is $v = \beta$.  If (\ref{kasner}), with $v<1$
and $\zeta \neq 0$, is substituted into the right hand side of
equations~(\ref{gowdyeq}) then the neglected terms are all
decaying for large enought $\tau$.  (The result in
\cite{grubisic93} takes into account corrections to all
orders and finds them to be decaying as $\tau \to \infty$.)
On the other hand, if (\ref{kasner}), with $v>1$
and $\zeta \neq 0$, is substituted into the right hand side of
equations~(\ref{gowdyeq}) then one of the terms in
equations~(\ref{gowdyeq}) that is absent from
equations~(\ref{gowdyinkasnerepoch}) eventually grows,
in agreement with the numerical simulations.

As the term grows, the one parameter family of Kasners stops being a good
approximation.  The numerical solution is next approximated well by a
``curvature bounce solution.''  The name refers to the solution's role as
the approximation of the transition from one Kasner Epoch to the next, in
which a term related to the spatial curvature grows and then decays again.
In a curvature bounce solution, $v \rightarrow |2 - v|$ \cite{berger97a},
in agreement with the numerical observations.  For a period of time, both
the one parameter family of Kasner solutions and the curvature bounce
solution approximate the Gowdy solution.  That is, the curvature bounce
solution approximates the latter part of a step, the transition and the
first part of the next step.

It turns out that the curvature bounce solutions are related to the
Kasner solutions.  Application of the transformation GE
to a solution of equations~(2) during a curvature bounce,
$(P,Q) \stackrel{GE}{\longrightarrow} (p,q)$, and then neglect
of small terms implies that $p$ and $q$ satisfy
equations~(\ref{gowdyinkasnerepoch}).  Therefore the expressions on
the right hand side of equations~(\ref{kasner}) can be used for $p$
and $q$.  The transformation GE is its own inverse, so transforming
$(p,q)$ back to $(P,Q)$, and defining $\pi_Q = -q_x$ gives
\begin{eqnarray}
\label{curvaturebounce}
P & = & -\ln \alpha - \ln(e^{-(\beta + 1) \, \tau}
+ \zeta^2 e^{(\beta-1) \, \tau} ) \nonumber \\
Q_x & = & 2 \alpha \beta \zeta \\
\pi_Q & = & - \xi_x - {e^{2 \beta \tau} \over
\alpha (1 + \zeta^2 e^{2 \beta \tau})} \left\{
{\alpha_x \zeta \over \alpha} - {2 \beta_x \zeta \tau
\over 1 + \zeta^2 e^{2 \beta \tau}} -
{\zeta_x (1 - \zeta^2 e^{2 \beta \tau}) \over
1 + \zeta^2 e^{2 \beta \tau}} \right\}
\nonumber
\end{eqnarray}
Since $(p,q)$ do not satisfy the full Gowdy equations, the 
integrability condition for $Q$ is not satisfied after the
transformation back.  If it were, then it would be the case
that $e^{-2 P} \pi_Q = Q_\tau$.  The conditions which must hold
so that $(p,q)$ approximately satisfy the full Gowdy
equations~(\ref{gowdyeq}) imply that the integrability condition
is approximately satisfied.  Upon substitution of
equations~(\ref{curvaturebounce}) into equations~(\ref{gowdyeq}),
one finds here too, that if $\beta<1$ and $\zeta \neq 0$,
the neglected terms are all decaying for large enough $\tau$,
but if $\beta>1$ and $\zeta \neq 0$, one of the
neglected terms in the equations~(\ref{gowdyeq}) eventually
grows.  In this case, the next stage of the evolution can
then be approximated by a one parameter family of Kasner
solutions~(\ref{kasner}) and the cycle continues.

As mentioned in section~\ref{gowdysec}, if no terms are neglected
after the transformation $(P,Q) \stackrel{GE}{\longrightarrow} (p,q)$,
then $(p,q)$ satisfy the Gowdy equations~(\ref{gowdyeq})
exactly.  In this case the result in \cite{grubisic93} can be applied
to $(p,q)$.  Transformation with GE to $(P,Q)$ then shows that
corrections to (\ref{curvaturebounce}) to all orders are decaying
as $\tau \to \infty$, if $\beta < 1$ and $\zeta \neq 0$.

The occurrence of spikes can be understood within this
approximation.  A ``downward pointing'' spike first appears during
a Kasner Epoch if $\zeta = 0$ at an isolated point.
An ``upward pointing'' spike first appears during a
curvature bounce, again if $\zeta = 0$ at an isolated point.  The
resulting shape of $P$ and $Q$ as given in equations~(\ref{kasner}),
and of $P$, $Q$ and $\pi_Q$ in equations~(\ref{curvaturebounce}),
qualitatively agree with the numerical observations.  If the
approximation remained a good one, this would imply that the
range of $\hat k$ at a spike is greater than obtained in
section~\ref{manufacture}.

To consider the magnitude of $\hat k$, let
$P_N = P_0 + \cdots + P_n$ and $Q_N = Q_0 + \cdots + Q_n$.  Let
$(P_0,Q_0)$ be given by equations~(\ref{kasner}) with $\zeta = 0$
at $x_0$.  At large $\tau$, $P_{0 \tau} \approx -\beta$ at $x_0$.
Let $P_{n+1}$ be obtained by substituting $P_{N+1}$ into
the left hand side of equations~(\ref{gowdyeq}) and $(P_N,Q_N)$ into
the right hand side and $Q_{n+1}$ be obtained in the analogous way.

At $x_0$, $P_{1 \tau \tau}$ is decaying for large enough $\tau$
if $0 \leq \beta < 1$ or for any $\beta \geq 0$ if the additional
condition, $\zeta_x = 0$ at $x_0$, is imposed.  However, if
$\beta > 1$ and $\zeta_x \neq 0$ at $x_0$, then $P_{1 \tau \tau}$
is growing exponentially for large $\tau$ and is strictly positive.
This suggests that $\hat k < -1$ will not occur in general at a spike.

Calculation of $P_{2 \tau \tau}$ shows that it is growing
exponentially at $x_0$ for large $\tau$ if $\beta > 2$, $\zeta_x = 0$
at $x_0$ and $\zeta_{x x} \neq 0$ there.  The dominating term is,
again, positive definite.  If $\zeta_{x x}$ does vanish
at $x_0$ and if $\beta < 3$, then $P_{2 \tau \tau}$ is decaying
for large enough $\tau$.  Inspection of the differential equation
satisfied by $P_{n \tau \tau}$ suggests that this pattern will
continue.  In the expression for $P_{n \tau \tau}$ there is
a term of the form $+(\partial^n_x \zeta)^2 e^{2(\beta - n) \tau}$.
If cancellation of this term does not occur, this suggests that
$\hat k < -n$ will not occur at a spike without conditions
on the spatial derivatives of $\zeta$ up to order n.

The conditions on the spatial derivatives of $\zeta$ can be
expressed as conditions on $\pi_Q$.  Let
$\pi_{Q 0} = e^{2 P_0} Q_{0 \tau}$.  From~(\ref{kasner})
it follows that $\pi_{Q 0} = - 2 \alpha \beta \zeta$.  If
$\zeta = 0$, then $\partial^m_x \pi_{Q 0} = 0$ is equivalent to
$\partial^m_x \zeta = 0$.  The analysis suggests that
$\hat k < -n$ does not occur at a spike unless
$\partial^j_x \pi_Q$ vanishes asymptotically at the spike,
for all integers $0 \leq j \leq n$.

Because of the relationship of equations~(\ref{curvaturebounce}) to
equations~(\ref{kasner}), the analysis of a spike which begins
when the evolution is approximated by a curvature bounce solution
can be obtained from the above by replacing the conditions on the
spatial derivatives of $\pi_Q$ with conditions on the spatial
derivatives of $Q_x$, and replacing $P_\tau \approx -\beta$ at large
$\tau$ with $P_\tau \approx 1 + \beta$.  So in this case the analysis
suggests that $\hat k > 1 + n$ does not occur unless $\partial^j_x Q_x$
vanishes asymptotically at the spike, for all integers $0 \leq j \leq n$.

\section{Further developments}\label{further}

In section~\ref{manufacture} we presented constructions of solutions of 
the Gowdy equations with spikes which are only local in $x$. Fortunately
it is not hard to see that these spikes can be grafted onto solutions
which are periodic in $x$ and which therefore define global Gowdy 
spacetimes. Suppose that we have a local solution as constructed above 
with a true spike at $x=x_{spike}$. Choose $\tau$ large enough and 
$|x-x_{spike}|$ small enough such that for a suitably chosen $k_*$ the
regions $R_1$ and $R_2$ defined in section \ref{manufacture} have the
desired properties. Let $\gamma_1$ and $\gamma_2$ be the unique curves
in the $(\tau,x)$-plane which are projections of null geodesics of the 
spacetime with $y$ and $z$ constant and which tend to $x_{spike}$ as
$\tau\to\infty$. Suppose $\tau$ chosen large enough so that the parts
of $\gamma_1$ and $\gamma_2$ with values of $\tau$ this large are 
contained in $R_2$. (To see that this is possible it is easier to think
in terms of $(t,x)$ coordinates, which make these curves into straight
lines.) Choose some $\tau_0$ so that the straight line $S$ given by 
$\tau=\tau_0$ intersects $R_1$. Let $S_2$ be the intersection of 
$R_2$ with $S$. The restriction of the initial data on $S$ for the given 
solution to a neighbourhood of $S_2$ already determines the features 
of the solution which are characteristic of true spikes. The given data
can be modified away from a neighbourhood of $S_2$ and extended to data 
which are periodic in $x$. The extended data determine a global Gowdy
spacetime which contains the original local spike. 

In \cite{chrusciel91} Chru\'sciel obtained a sufficient criterion for
the blow-up of the Kretschmann scalar near a point of the singularity
in a Gowdy spacetime. This is his equation (3.4.5) which implies that
the leading order behaviour of the curvature is as in the corresponding
Kasner solution.  It is interesting
to see that this criterion is satisfied for the generic true spikes 
constructed above. This suggests that the techniques he uses to prove 
cosmic censorship in Gowdy spacetimes may admit extensions to wider 
classes, or even to the general case. Suppose that we have a solution
of the Gowdy equations with a true spike at $x=0$ of the kind
constructed in section \ref{manufacture}. Suppose further that the
genericity condition $X_{0x}(0)\ne 0$ is satisfied. The criterion
of \cite{chrusciel91}, reexpressed in our variables, is
\begin{equation}
\label{criterion}
\int_{t_0}^0\int_{-t}^t (P_x^2+e^{2P}Q_x^2) dx dt < \infty
\end{equation}    
The region of integration is entirely contained in the region $R_2$ when
$t_0$ is close to zero. Using the asymptotic expansion of the solution 
which is valid in that region shows that, up to terms of lower order,
the integral to be computed is proportional to
\begin{equation}
\int_{t_0}^0\int_{-t}^t t^{-2\hat k} x^2 dx dt 
\end{equation}
The integral is finite when $\hat k<2$, which is a condition which 
naturally arises from our construction.

Next we produce spikes which have asymptotic velocity which is
greater than that of the spikes produced in
section~\ref{manufacture}, and which are consistent with the
conditions suggested by the analysis in section~\ref{MCP}.
An inductive argument shows that successive applications
of the transformations (\ref{inversion}) and (\ref{gowernst})
produce spikes with $|\hat k|$ arbitrarily large.
To begin the inductive argument, let $i$ be a nonnegative integer.
Given a Gowdy solution labeled $(P_{2i},Q_{2i})$,
let $(P_{2i},Q_{2i}) \stackrel{I}{\longrightarrow} (P_{2i+1},Q_{2i+1})$
define $(P_{2i+1},Q_{2i+1})$.
Let $(P_{2i+1},Q_{2i+1}) \stackrel{GE}{\longrightarrow}
(P_{2i+2},Q_{2i+2})$ define $(P_{2i+2},Q_{2i+2})$.  Then
\begin{eqnarray}
P_{2i+1} & = & P_{2i} + \ln( Q_{2i}^2 + e^{-2P_{2i}}) \nonumber \\
\label{induct1}
Q_{2i+1} & = & {Q_{2i} \over Q_{2i}^2 + e^{-2P_{2i}}} \\
\pi_{Q_{2i+1}} & = & - e^{2 P_{2i}} Q^2_{2i} \partial_\tau Q_{2i} +
\partial_\tau Q_{2i} + 2 \, (\partial_\tau P_{2i}) \, Q_{2i}. \nonumber
\end{eqnarray}
And
\begin{eqnarray}
P_{2i+2} & = & \tau - P_{2i} - \ln( Q_{2i}^2 + e^{-2P_{2i}}) \nonumber \\
\label{induct2}
\partial_\tau Q_{2i+2} & = & 
e^{2 ( P_{2i} - \tau )} Q_{2i}^2 \, \partial_x Q_{2i} -
e^{- 2 \tau} ( \partial_x Q_{2i} + 2 \, (\partial_x P_{2i}) \, Q_{2i}) \\
\partial_x Q_{2i+2} & = & e^{2 P_{2i} } Q_{2i}^2 \, \partial_\tau Q_{2i} -
\partial_\tau Q_{2i} - 2 \, (\partial_\tau P_{2i}) Q_{2i}. \nonumber
\end{eqnarray}
Let $(P_0,Q_0)$ be as in section~\ref{manufacture}.  Then the
following four properties hold for $i=0$.

\vspace{10pt}
1) $\displaystyle\lim_{\tau \rightarrow \infty}\partial^j_x Q_{2i}=0$
at $x_{spike}$
for all non-negative
integers $j$, such that $j \leq i$.

2) $i < \displaystyle\lim_{\tau \rightarrow \infty}\partial_\tau P_{2i} < i+1$
at $x_{spike}$, while
$0 < \displaystyle\lim_{\tau \rightarrow \infty}\partial_\tau P_{2i}
< 1$ elsewhere in a neighbourhood of $x_{spike}$.

3) $\displaystyle\lim_{\tau \rightarrow \infty}
\partial^j_x \pi_{Q_{2i+1}} = 0$ at $x_{spike}$
for all nonnegative integers $j$ such that $j \leq i$.

4)$-(i+1) < \lim_{\tau \rightarrow \infty}\partial_\tau P_{2i+1}<-i$
at $x_{spike}$, while
$0 < \lim_{\tau \rightarrow \infty}\partial_\tau P_{2i+1}<1$
elsewhere in a neighbourhood of $x_{spike}$.

\vspace{10pt}
Now assume that these four properties hold for $i$ equal to some
nonnegative integer, $n$.  Then from equations (\ref{induct1}) and
(\ref{induct2}), and if the constant of integration for $Q_{2n+2}$
is set such that $\lim_{\tau \rightarrow \infty} Q_{2n+2} = 0$
at $x_{spike}$, it follows that the four properties hold for $i=n+1$.
So by induction, these four properties hold for any nonnegative
integer, $i$, with the constant of integration thus determined for
each application of the GE transformation.

The spikes just produced are consistent with the following conjecture,
as is the analysis in section~\ref{MCP}.  Given some non-negative
integer $n$, if $-n > \hat k_{spike} > - (n+1)$, then at $x_{spike}$,
$\lim_{\tau \rightarrow \infty} \partial^m_x \pi_Q = 0$ for all
integers, $0 \leq m \leq n$. And given some positive integer $n$, if
$n < \hat k_{spike} < (n+1)$ then at $x_{spike}$,
$\lim_{\tau \rightarrow \infty} \partial^m_x Q = 0$
for all integers, $1 \leq m \leq n$.

Note that $(P_{2i+1},Q_{2i+1})$ is equivalent to $(P_{2i},Q_{2i})$
since they are related by inversion.  Thus, to determine whether
the spikes just produced are true or false we need only examine
the curvature of $(P_{2i},Q_{2i})$.  If criterion~(\ref{criterion})
is satisfied, the curvature of $(P_{2i},Q_{2i})$ near the singularity
is as in the corresponding Kasner spacetimes.  Therefore the discussion
at the end of section~\ref{manufacture} applies with $0<\hat k<1$
for $x\ne x_{spike}$ and $i < \hat k(x_{spike}) < i+1$.  It
follows that $(P_{n},Q_{n})$ represents a true spike if $n>1$ and
criterion~(\ref{criterion}) is satisfied.

That criterion~(\ref{criterion}) is satisfied can be checked
by obtaining expressions for $(P_{2i},Q_{2i})$ on the region of
integration.  At $x_{spike}$ these are, for small enough $t_0$,
\begin{equation}
\label{highvexpansion}
P_{2i} = (i + k) \tau + O(1) \hspace{40pt}
Q_{2i} = X_{2i} + o(e^{-c \tau})
\end{equation}
for some $c>0$, and with $\partial^j_x X_{2i}$ vanishing at
$x_{spike}$ for all integers, $1 \leq j \leq i$.  These
expressions follow from arguments similar to those concerning
the expressions for $(P_2,Q_2)$ on $R_2$.  The quantities which
must be estimated on the region of integration are of the
form $e^{(m + k) \tau} |X_{2m}|$, with $\partial^j_x X_{2m}$
vanishing at $x_{spike}$
for $1 \leq j \leq m$.  For $x$ close enough but not equal
to $x_{spike}$, $P_{2i} = \hat k \tau + O(1)$ on the region of
integration in criterion~(\ref{criterion}) for small enough $t_0$,
with $0 < \hat k < 1$. The expression for $Q_{2 i}$ is that given
in (\ref{highvexpansion}).  It follows that criterion~(\ref{criterion})
is satisfied in a neighbourhood of $x_{spike}$.

Because of the restrictive conditions which are satisfied at
these higher velocity spikes, it is unlikely for them to
have occurred in numerical simulations, since initial
data for numerical simulations has not been specially
chosen to obtain them.

If we look beyond Gowdy spacetimes, the point of view of BKL and the method 
of consistent potentials suggest that there is a close association between
spikes and oscillations in time. According to the picture of BKL general
solutions of the vacuum Einstein equations should show behaviour near the
singularity similar to that of the mixmaster solution and, in particular,
an infinite number of oscillations. This could lead to the generation of
densely distributed inhomogeneous structure, which even calls the 
consistency of the BKL picture into question (cf. \cite{belinskii92}.)
Issues of this type are beyond the reach of the rigorous mathematical
techniques available at this time. However we can address one interesting
issue. In \cite{andersson00} it was proved, following a suggestion of
Belinskii and Khalatnikov \cite{belinskii73} that coupling the Einstein
equations to a scalar field can lead to mixmaster oscillations being
suppressed. (See also the results on the homogeneous case in
\cite{ringstrom00b}.) One might then ask if the presence of a scalar field
could suppress the occurrence of (true) spikes. We will show that this
is not the case. In fact this is apparent within the BKL approach. The
scalar field forbids the occurrence of an infinite number of oscillations
and hence the occurrence of an unlimited number of spikes. However it
does not forbid finitely many oscillations and finitely many spikes.

The occurrence of true spikes in solutions of the Einstein-scalar field
system will now be shown rigorously. The idea is to take a solution
of the Gowdy equations with a true spike and produce from it a solution
of the Einstein-scalar field equations with Gowdy symmetry which also
contains a spike. The possibility of doing this is based on the structure 
of the field equations in this case. The metric can be written in the 
same form as in the vacuum case. The field equations look very similar to 
the equations in the vacuum case. The equations for $P$ and $Q$ are 
unchanged. This is because they come from the projection of the Ricci
tensor onto the orbits of the group action. Since the Ricci tensor
is proportional to the tensor product of the gradient of the scalar field
with itself, and this gradient has vanishing projection onto the orbits,
the corresponding components of the Einstein equations are unchanged. The 
scalar field satisfies the equation
\begin{equation}
\phi_{\tau\tau}=e^{-2\tau}\phi_{xx}
\end{equation} 
Note that this is the same equation as satisfied by $P$ in the case that 
$Q$ is zero. Thus there is a strong connection to polarized Gowdy
solutions. In particular, we understand the asymptotics of solutions
of the equation for $\phi$ completely. The equations for $\lambda$ are 
modified by adding to the right hand sides contributions from $\phi$
formally identical to those due to $P$. Let us start with a solution
of the Gowdy equation with a true spike and take any non-vanishing
solution of the equation for $\phi$. We can then construct a solution
of the Einstein-scalar field equations as just indicated. The function
$P$ has not been changed and so still has a spike. The calculation for
the curvature is essentially the same as in the vacuum case. In the 
calculation for the general case the leading order is given by the 
Bianchi I case. The function $\phi$ is asymptotically of the form 
$A\ln t+B$. The only difference in the computation of the Kretschmann
scalar is that the Kasner exponents are replaced by the generalized
Kasner exponents satisfying the equation $p_1^2+p_2^2+p_3^2=1-16\pi A^2$. 
If $A$ is small then the limits of the generalized Kasner exponents of the 
solutions with scalar field at the singularity will differ only a little 
from the limits of the Kasner exponents of the original vacuum solution.
It follows that the addition of a scalar field does not always remove
the discontinuous limiting behaviour of the Kretschmann scalar.  

The fact that we were able to obtain so much information about spikes
in this paper was based on a trick (the use of the inversion and
Gowdy-to-Ernst transformation). Spikes can be expected to occur in
much more general spacetimes but this trick cannot be expected to
extend very far. The method of consistent potentials is not limited in 
the same way and it would be very desirable to transform that technique 
into a tool which can be used to prove theorems. Having one case, namely
the Gowdy case, which is understood in detail may provide valuable 
guidance as to how to do this.

\vskip 10pt\noindent
{\it Acknowledgements} We thank Marc Mars, Vincent Moncrief and 
John Stewart for enlightening suggestions concerning the subject of this 
paper.  MW thanks Beverly Berger and James Isenberg for discussions
concerning the analysis in section~\ref{MCP}.

\end{document}